\documentclass[%
 twocolumn,
 amsmath,amssymb,
 aps,
]{revtex4-2}

\usepackage{color}

\usepackage[dvipsnames]{xcolor}
\usepackage{graphicx}% Include figure files
\usepackage{dcolumn}% Align table columns on decimal point
\usepackage{bm}% bold math
\begin{document}

\preprint{APS/123-QED}

\title{Dual-species Bose-Einstein condensates of $^{7}$Li and $^{133}$Cs}% Force line breaks with \\

\author{Y.-D. Chen}
\author{W.-X. Li}
\author{Y.-T. Sun}
\author{Q.-C. Chen}
\author{P.-Y. Chang}
\author{S.~Tung}
\email{stung@phys.nthu.edu.tw}
\affiliation{%
 Department of Physics, National Tsing Hua University, Hsinchu 30013, Taiwan\\
  and Center for Quantum Technology, Hsinchu 30013, Taiwan
}%

\date{\today}% It is always \today, today,
\begin{abstract}

We report the creation of dual-species Bose-Einstein condensates (BECs) of $^{7}$Li and $^{133}$Cs. These BECs are formed in a bichromatic optical dipole trap created with 1550- and 780-nm laser beams. During the production process, an external magnetic field of 886~G is applied to adjust the scattering lengths to $a_{\rm{Cs}} = 123a_0$, $a_{\rm{Li}} = 484a_0$, and $a_{\rm{LiCs}} = 248a_0$. These scattering lengths allow for efficient evaporation and sympathetic cooling. The dual-species BECs are typically produced with  $1.5\times 10^4$ Cs atoms and $6.0\times 10^3$ Li atoms. This quantum degenerate mixture of Li and Cs provides an ideal platform for exploring phenomena such as polarons and Efimov trimers, as well as for creating ground-state LiCs molecules.

\end{abstract}

\maketitle

%----------------------------------------------------------------------------------

\section{introduction}
Quantum gas mixtures provide an ideal platform for exploring paradigms of many-body and few-body physics. These mixtures can be created with atoms of the same species in different spin states \cite{PhysRevLett.78.586, PhysRevLett.85.2413} or isotopes \cite{Science.291.2570, PhysRevLett.87.080403,PhysRevA 79.021601}, or of different species \cite{PhysRevLett.88.160401, PhysRevLett.89.150403, PhysRevLett.100.010401, PhysRevA.84.011603, PhysRevA.85.051602, JPhysB.49.015302, PhysRevLett.119.233401, PhysRevA.104.033302}. Two-species mixtures present special interests due to their capacities for large mass imbalance and species-specific optical manipulation. These  features make them suitable for simulating fascinating phenomena such as polarons \cite{PhysRevLett.102.230402, nature11065, PhysRevLett.117.055301, PhysRevLett.117.055302}, quantum droplets \cite{PhysRevLett.115.155302, Science.359.301, PhysRevLett.120.235301, PhysRevLett.120.135301}, and Efimov trimers \cite{Nature.440.315, PhysRevLett.103.043201, NaturePhys.5.227, NaturePhys.5.586, RepProgPhys.80.056001,RevModPhys.89.035006}. Alkali-metal mixtures are particularly attractive because of their readily accessible magnetic Feshbach resonances \cite{RevModPhys.82.1225}. These resonances not only offer precise control over inter- and intraspecies interactions but also play a vital role in the creation of ground-state polar molecules \cite{Science322.231, PhysRevLett.113.205301, PhysRevLett.113.255301, PhysRevLett.114.205302, PhysRevLett.116.205303, PhysRevLett.125.083401, PhysRevLett.130.113002}.

Alkali-metal mixtures composed of lithium (Li) and cesium (Cs) exhibit intriguing properties. Depending on the isotope of Li used, Li-Cs mixtures can exhibit either Bose-Fermi ($^{133}$Cs-\,$^{6}$Li) or Bose-Bose ($^{133}$Cs-\,$^{7}$Li) quantum statistical properties. The distinct optical polarizabilities of the two species enable easy implementation of species-specific optical potentials \cite{PhysRevA.73.022505, doi.org/10.1016/S1049-250X(08)60186-X}. Additionally, the mass imbalance is substantial, $m_{\rm{Cs}}/m_{\rm{Li}}\approx20$, and the ground-state LiCs molecules possess a large permanent dipole moment of 5.5 D \cite{J.Chem.Phys.112.204302, PhysRevA.82.032503}.

Our current focus is on the $^7$Li-\,$^{133}$Cs mixtures. Despite only a 10\% change in mass imbalance compared to $^6$Li-\,$^{133}$Cs mixtures, $^7$Li-\,$^{133}$Cs mixtures allow for a completely different exploration due to their distinct quantum statistics. For instance, the mixtures present unique opportunities to contrast $^7$Li-\,$^7$Li-\,$^{133}$Cs and $^7$Li-\,$^{133}$Cs-\,$^{133}$Cs Efimov trimers. Also, the mixtures permit studies of Bose polarons created with both heavy and light impurities.

Earlier studies of the Li-Cs mixtures include the investigation of the collisional properties of ultracold $^{7}$Li and $^{133}$Cs atoms \cite{Eur.Phys.J.D.7.331, Appl.Phys.B.73.791, PhysRevLett.88.253001, PhysRevA.70.062712}, as well as the creation of ground-state $^{7}$Li$^{133}$Cs molecules through photoassociation \cite{PhysRevLett.101.133004}. Later research has used $^{6}$Li-\,$^{133}$Cs mixtures to study  Feshbach resonances \cite{PhysRevA.87.010701, PhysRevA.87.010702, PhysRevA.90.012710} and explore Efimov physics \cite{PhysRevLett.112.250404, PhysRevLett.113.240402, PhysRevA.93.022707, NaturPhysics.13.731}. Recently, a degenerate Fermi gas of $^{6}$Li trapped in a Cs Bose-Einstein condensate (BEC) has been realized \cite{PhysRevLett.119.233401}, and $^{6}$Li -mediated interactions between Cs atoms have been observed in that system \cite{Nature568.61}. Last year, we reported the identification of ten interspecies Feshbach resonances in the  $^{7}$Li-\,$^{133}$Cs mixtures  \cite{PhysRevA.106.023317}.

In this paper, we report on the creation of a quantum gas mixture, consisting of overlapping BECs of $^{7}$Li and $^{133}$Cs. In the production process, we use a sequential loading method to prepare ultracold Li and Cs atoms in separate optical dipole traps (ODTs). Subsequently, Li is transferred to the Cs trap, followed by evaporation and sympathetic cooling in the presence of an external magnetic field of 886~G. This process results in the formation of dual-species BECs composed of $6.0\times10^3$ Li atoms in the $\vert F=1, m_F=0\rangle$ state and $1.5\times10^4$ Cs atoms in the $\vert F=3, m_F=3\rangle$ state.

The rest of this paper is organized as follows: Section~\ref{mixture_preparation} outlines the experimental methods for preparing ultracold Li and Cs atoms. Section~\ref{ODTs} provides details on the far-detuned optical dipole traps in this work and discusses their roles in creating dual-species BECs. Section~\ref{Evap_and_Merging} describes our technique for merging and evaporating Li and Cs atoms. Section~\ref{sympathetic_cooling} covers the sympathetic cooling and the final evaporation procedure. We report our observation of dual-species Li-Cs BECs in Sec.\ref{BEC}, and finally, we summarize our paper in Sec.\ref{conclusion}.

\section{ultracold lithium and cesium atoms} \label{mixture_preparation}

\begin{figure}[t]
	\centering
	\includegraphics[width=0.5\textwidth]{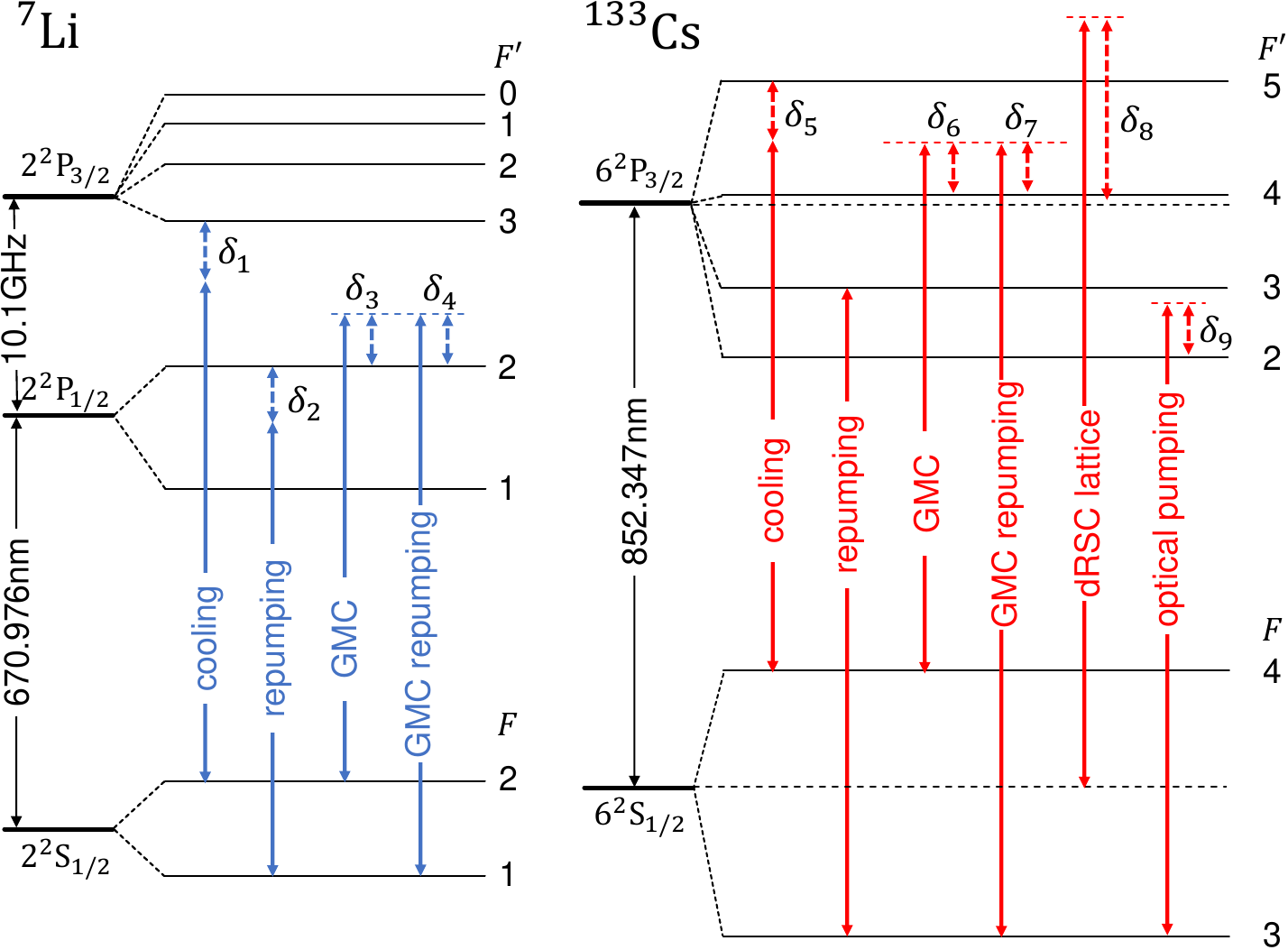}
	\caption{Diagram of the relevant optical transitions used in our experiment. The transitions are represented by the vertical arrows, with the detunings given by $\delta_1 = -31$ MHz (-5.3$\Gamma$), $\delta_2 = -16$ MHz (-2.7$\Gamma$), $\delta_3 = \delta_4 = 19$ MHz (3.2$\Gamma$), $\delta_5 = -12$ MHz (-2.3$\Gamma$), $\delta_6 = \delta_7 = 28$ MHz (5.4$\Gamma$), $\delta_8 = 16$ GHz, and $\delta_9 = 10$ MHz (1.9$\Gamma$). These detunings are relative to the atomic transition frequencies depicted, with values in parentheses indicating the detuning in terms of the respective transition linewidth. Consequently, the specific values of $\Gamma$ can vary.}
	\vspace{-4mm}
	\label{fig:energy_level}
\end{figure}

Our experimental setup utilizes the Li-Cs slow beam detailed in our previous work \cite{PhysRevA.103.023102} along with a dual-species magneto-optical trap (MOT) to prepare ultracold Li and Cs atoms. For Li, the MOT beams consist of cooling light, red detuned (-31~MHz or -5.3~$\Gamma$) to the $2S_{1/2}(F = 2) \rightarrow 2P_{3/2}(F' = 3)$ transition, and repumping light red detuned (-16~MHz or -2.7~$\Gamma$) to the $2S_{1/2}(F = 1) \rightarrow 2P_{1/2}(F' = 2)$ transition. For the Cs MOT beams, we use the $F = 4 \rightarrow F'=5$ transition with a detuning of -12~MHz (-2.3 $\Gamma$) for cooling and the $F = 3 \rightarrow F'=3$ transition for repumping, both in the Cs D2 line. Here, $\Gamma$ represents the natural linewidth of the respective transition. In Fig.~\ref{fig:energy_level}, we summarize the relevant optical transitions for laser cooling of Li and Cs. 

The experiment procedure begins with Li MOT loading, which collects approximately $3 \times 10^7$ Li atoms in 30 s. With Li temperatures in the MOT reaching as high as 650~$\mu$K, achieving efficient ODT loading poses an acute challenge. To overcome the difficulty, we implement D1 gray molasses cooling (GMC) \cite{PhysRevA.87.063411, PhysRevA.99.053604} on Li after a 20-ms MOT compression. The gray molasses is formed by adapting the existing MOT beams; we extinguish the cooling light and shift the frequency of the repumping laser +19 MHz (+3.2~$\Gamma$) above the $2S_{1/2}(F = 2) \rightarrow 2P_{1/2}(F' = 2)$ transition. Simultaneously, we phase modulate the beam to generate a sideband that couples the $2S_{1/2}(F = 1) \rightarrow 2P_{1/2}(F' = 2)$ transition, with the same detuning. During GMC, the four horizontal beams have an average intensity of 22 mW/cm$^2$, while the two vertical beams have an intensity of 50 mW/cm$^2$. After implementing the molasses cooling for 5 ms, 60\% of the Li atoms, corresponding to $N_{\rm{Li}} = 1.8\times 10^7$, are cooled to approximately 25~$\mu$K, which allows for more effective ODT loading of the atoms.

\begin{figure}[t]
	\centering
	\includegraphics[width=0.5\textwidth]{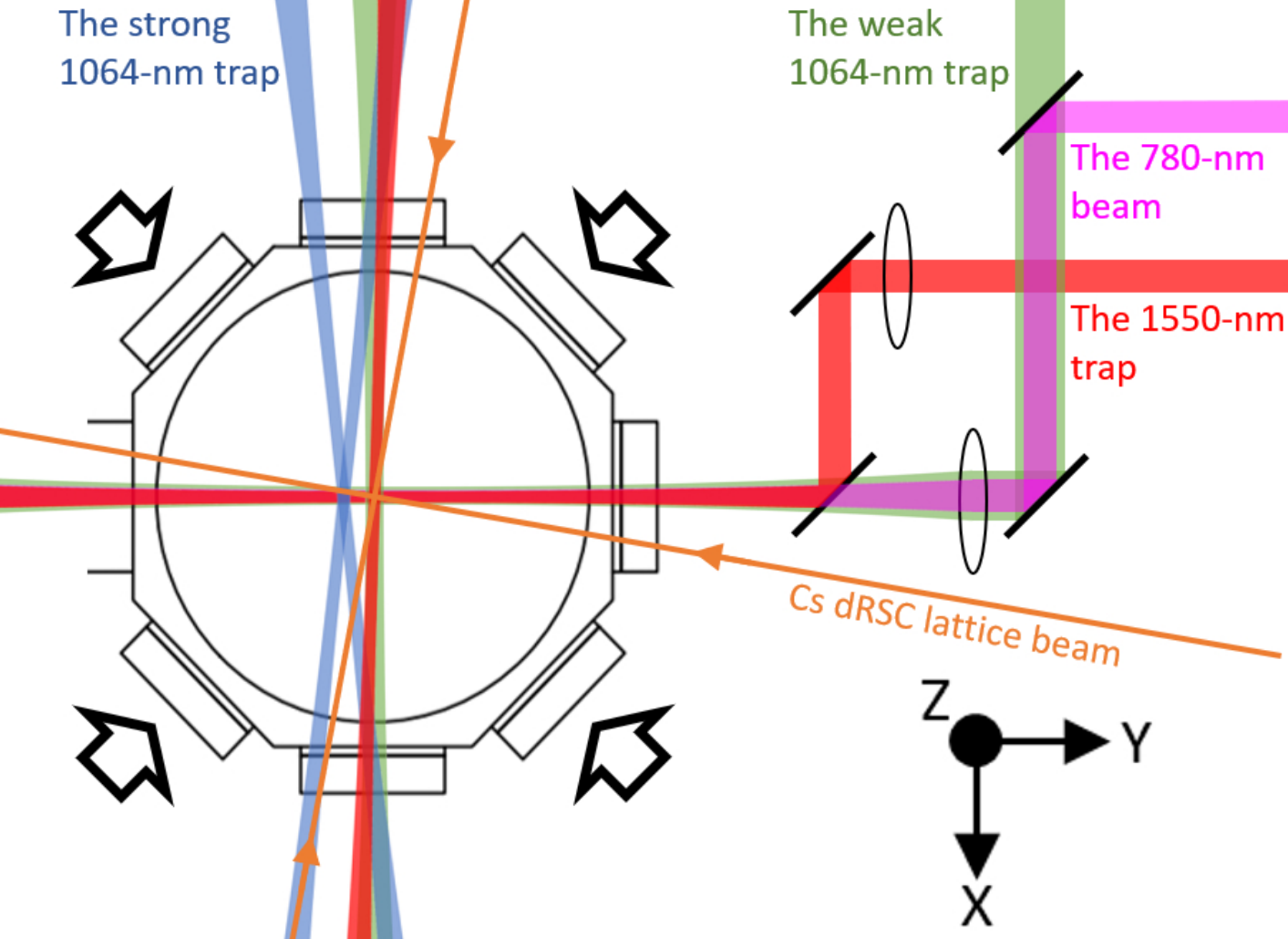}
	\caption{The optical layout for the dipole traps, Cs dRSC lattice, and MOTs (not to scale). Three dipole traps are utilized in this work: (1) the strong 1064-nm trap (blue), referred to as the Li ODT due to its role in Li loading; (2) the weak 1064-nm trap (green); and (3) the 1550-nm trap (red). The last two traps together form the Cs ODT. An additional 780-nm laser beam (magenta), overlapping with the $y$ beam of the 1550-nm trap, serves not only to adjust the relative trap depths for Li and Cs but also to facilitate the Li transfer to the 1550-nm trap. The Li ODT is initially offset by 350 $\mu$m in the $y$ direction relative to the center of the 1550-nm trap. The figure also depicts three dRSC lattice beams (orange). The remaining dRSC beams, not shown in this figure, are aimed towards the Cs atoms along the $z$ axis. This figure illustrates the directions of the MOT beams by four large arrows and denotes the direction of gravity as $-z$.
}
\vspace{-4mm}
	\label{fig:optical layout}
\end{figure}

For Cs, we load the Cs MOT for 5 s, capturing $9\times10^7$ atoms from the slow beam source. After a 40-ms compression phase for the MOT, we execute two additional laser cooling steps: GMC and degenerate Raman sideband cooling (dRSC) \cite{PhysRevLett.81.5768, Phys.Rev.Lett.84.439, PhysRevLett.85.724, PhysRevA.95.033412}. The Cs GMC, like that of Li, is implemented with the existing MOT setup but utilizes the transitions in the Cs D2 line ($6S_{1/2} \rightarrow 6P_{3/2}$) \cite{PhysRevA.98.033419}. The molasses is produced with a total optical power of 24 mW; the horizontal beams possess an average intensity of 2 mW/cm$^2$, whereas the vertical beams display an intensity of 4.4 mW/cm$^2$. The majority of this power drives the cooling transition ($F = 4 \rightarrow F'=4$), with 4$\%$ used for repumping ($F = 3 \rightarrow F' = 4$). Following a 4-ms molasses cooling, the Cs temperature is reduced to 6~$\mu$K. Afterwards, the Cs atoms are transferred to a three-dimensional (3D) optical lattice for dRSC. The arrangement of the dRSC lattice beams and their polarizations is identical to those described in Ref.~\cite{PhysRevA.95.033412}. The lattice is formed by four linearly polarized laser beams, two counter-propagating along the $x$ axis and two along the $y$ and $z$ axes (see Fig.\ref{fig:optical layout}), with a frequency 16 GHz above the $6S_{1/2} \rightarrow 6P_{3/2}$ transition. Optical pumping and repumping beams for dRSC are both sent in from the $y$ direction. We use the $6S_{1/2}(F = 3) \rightarrow 6P_{3/2}(F' = 2)$ transition for optical pumping and the $6S_{1/2}(F = 4) \rightarrow 6P_{3/2}(F' = 3)$ transition for repumping. After completing dRSC, we obtain $5\times10^7$ Cs atoms in the $\vert F = 3, m_F=3\rangle$ state with a temperature slightly below 1$\mu$K after adiabatic release.

\section{Far-detuned Optical dipole traps}\label{ODTs}
In this experiment, we employ three optical dipole traps; two use light derived from a 200-W 1064-nm fiber laser, while the third uses light from a 20-W single-frequency 1550-nm fiber laser. The traps have been designed with specific depths and volumes to facilitate efficient evaporation. To aid in visualizing the trap arrangement, see Fig.~\ref{fig:optical layout} for a schematic of the experimental setup.

The first (strong) 1064-nm trap, referred to as the Li ODT, is created by crossing two focused laser beams at an angle of 10$^\circ$ in the region where the Li atoms reside. The two beams have waists of 150 $\mu$m (vertical) and 50 $\mu$m (horizontal), and each beam has a maximum optical power of 55 W, resulting in a trap depth of $\approx$0.6~mK. Furthermore, the switching and pointing of the Li ODT beams are controlled by an acousto-optic deflector (AOD). By changing the frequency of the rf source for the AOD, we can shift the trap position as needed. To ensure that the trap does not lose the crossing point during vertical movement, we take care to make the two beam paths as symmetric as possible. Also, during alignment, we verify that the two beams stay in the same plane with the axis of the lens and hit the lens perpendicularly with equal distances from the lens center.

The second (weak) 1064-nm trap is also created by crossing two focused laser beams, but at an angle of $\approx$90$^{\circ}$. Both beams have waists of 270 $\mu$m (vertical) and 380 $\mu$m (horizontal), and each beam has a maximum power of 2.9 W. The trap features a large trapping volume that covers a significant portion of the Cs cloud after dRSC. However, the trap alone is not strong enough to hold Cs against gravity. To counteract the gravitational pull, a magnetic-field gradient is applied to levitate Cs. The Cs atoms in the $\vert 3, 3\rangle$ state require a gradient of 31~G/cm to nullify the gravity  effect. Meanwhile, the introduction of this magnetic-field gradient reduces the trap depth of the Li ODT by approximately 6\%.

The 1550-nm trap, also a crossed ODT, is constructed using laser beams with a horizontal waist of 150 $\mu$m and a vertical waist of 30 $\mu$m. This trap, often referred to as a dimple trap, works in combination with the weak 1064-nm trap to form the Cs ODT. The weak 1064-nm trap, with its large trapping volume, facilitates the loading of a larger number of Cs atoms. In contrast, the 1550-nm trap, with its compact trapping region at the center, enhances Cs density and aids in more efficient evaporative cooling. Additionally, the tight confinement of the 1550-nm trap improves the overlap between Li and Cs.

\begin{figure*}
	\includegraphics[width=1.05\textwidth]{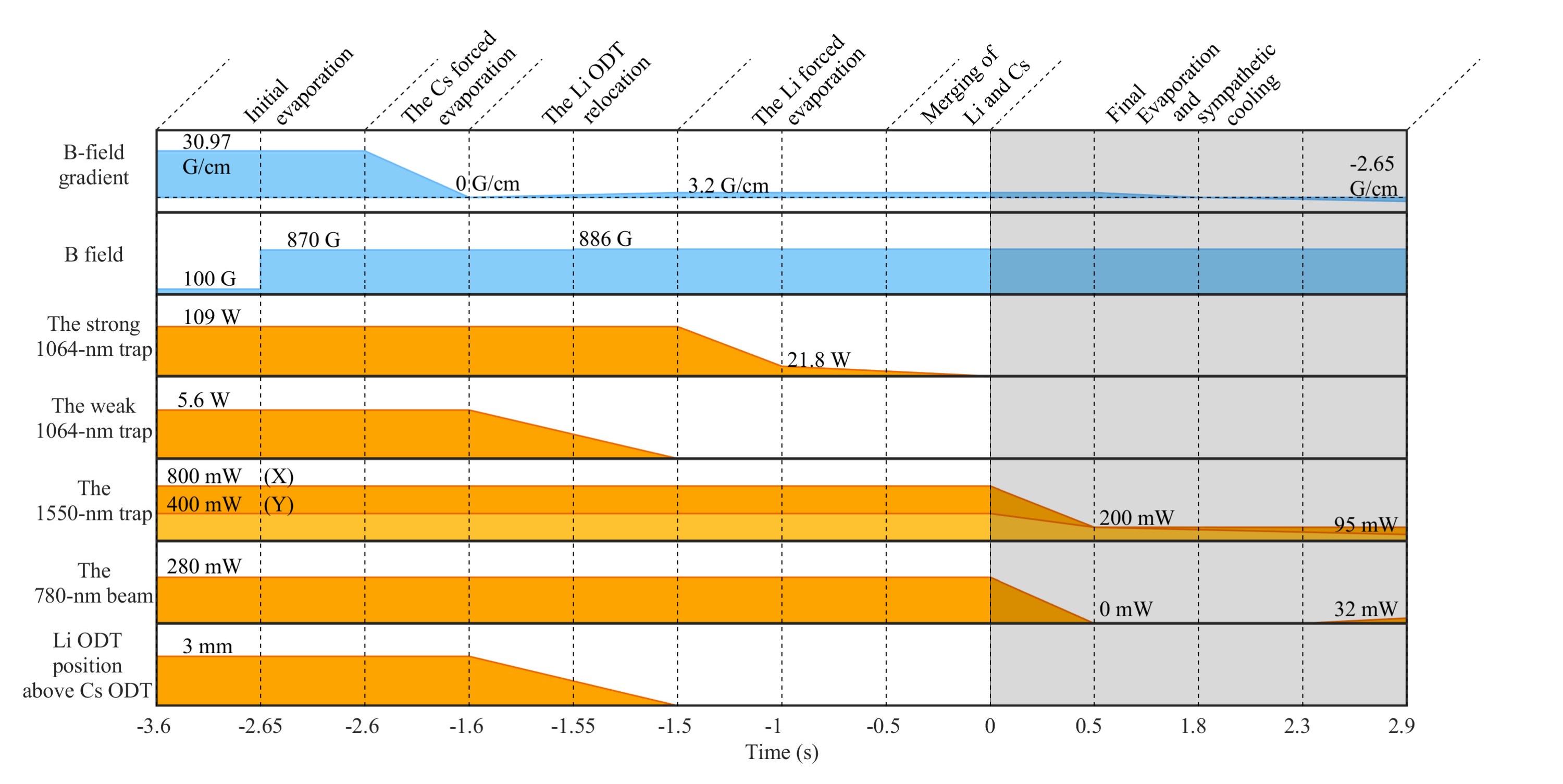}
	\caption{Timing sequence for evaporation and sympathetic cooling of Li and Cs. The sequence is divided into two parts: the pre-merge stage (white background) and the post-merge stage (gray background). During the pre-merge stage, evaporative cooling is performed simultaneously on Li and Cs in separate traps. Afterwards, Li is transferred to the 1550-nm trap where Cs resides. In the post-merge stage, the evaporation of the two species is carried out by weakening the 1550-nm trap and gradually ramping off the 780-nm beam. In the last phase of evaporation, we reactivate the 780-nm beam to enhance the overlap of the two species.}
	\vspace{-2.5mm}
	\label{fig:expt_proc}
\end{figure*}

We employ a sequential loading approach to introduce the two species into their respective dipole traps, namely the Li ODT and the Cs ODT. This strategy addresses the challenge of accommodating the distinct trap requirements for optimal evaporation and ensures the correct loading of atoms into their corresponding traps. Following is the detailed overview of this strategy: first, we load the Li MOT. Then, the Li atoms are cooled using GMC and subsequently loaded into the Li ODT. Notably, the Li ODT is initiated during the Li MOT loading. Next, the Li ODT is shifted upwards by approximately 3 mm, effectively moving the Li atoms along with it. During the shifts of the trap, we do not observe any substantial Li losses beyond the expected number of fluctuations. This displacement establishes an appropriate spatial separation between the two species, mitigating undesired loading of Cs into the Li ODT and preventing Li-Cs light assisted collision losses. Afterwards, the Cs MOT is loaded, followed by the application of Cs GMC and dRSC. Once the Cs atoms have been sufficiently cooled, they are loaded into the Cs ODT. The Cs ODT is activated during the Cs MOT loading. Finally, both Li and Cs atoms are trapped in their respective ODTs for evaporation. 

Following our sequential loading strategy, we are able to load $1.5\times10^6$ Li atoms into the Li ODT, yielding a loading efficiency of approximately 8\%. In the case of Cs, we manage to load 20\% of the Cs atoms from the dRSC lattice into the Cs ODT, which translates to roughly $1 \times 10^7$ Cs atoms. At the completion of the Cs loading procedure, we observe a 70\% reduction in the Li atom number in the Li ODT, mainly due to processes such as evaporation and spin-exchange collisions. Consequently, the Li ODT population decreases to approximately $4.7\times10^5$ Li atoms.

\section{Evaporation and Merging}\label{Evap_and_Merging}
Merging Li and Cs into the same ODT while preserving their phase-space densities can be challenging, especially when the two species have significantly different temperatures. To overcome this issue, we execute evaporative cooling for each species in their separate traps, thereby reducing the temperature difference between them.

The Cs atoms undergo evaporation in the Cs ODT following their transfer from the dRSC lattice. Initially, we hold the Cs atoms in the Cs ODT for 1 s with a bias magnetic field of 100 G and a field gradient of 31 G/cm. We empirically select the bias magnetic field to enhance the evaporation efficiency. We perform Cs forced evaporation by abruptly changing the field to 870 G and gradually reducing the magnetic-field gradient (see Fig.\ref{fig:expt_proc}). As the field gradient diminishes, the weak 1064-nm trap slowly opens, allowing the hotter atoms to evaporate. We select 870~G as the bias field value to ensure that the Li atoms, situated 3 mm above Cs, avoid the Li Feshbach resonance at 894~G \cite{CRPhys.12.4}. 

As the magnetic-field gradient reduces during Cs evaporation, we align the Li ODT with the Cs ODT in the same x-y plane. Subsequently, the bias field is increased to 886~G to optimize the evaporation efficiency for both Li and Cs.  Following the field change, we gradually reduce the power of the Li ODT to perform Li forced evaporation.

After completing separate evaporative cooling on Li and Cs, we proceed to combine them in the 1550-nm trap. When the Li ODT is adjusted to the same height as Cs, one of the 1550-nm beams passes through its center. Subsequently, we turn off the Li ODT and transfer the Li atoms to the Cs trap. However, at this stage, the 1550-nm trap lacks sufficient depth to effectively trap Li, leading to a loss of approximately 80$\%$ of the Li atoms. While increasing the trap depth of the 1550-nm ODT could improve the Li transfer, it would also raise the Cs temperature, leading to excessive Li losses during thermalization with Cs. To address this issue, we introduce an additional dipole beam with a wavelength of 780 nm.

The 780-nm beam, possessing a horizontal waist of 130 $\mu$m and a vertical waist of 28 $\mu$m, overlaps with the 1550-nm dipole beam along the y axis and passes through the centers of the Cs and Li ODTs, as shown in Fig.\ref{fig:optical layout}. This beam generates an optical potential that is repulsive for Cs and attractive for Li.  By adjusting the optical power of the 780-nm beam, we can adjust the relative trap depths of the two species. This is crucial for the next evaporation stage and sympathetic cooling of the two species. Similar approaches have been utilized in other mixture experiments such as $^{6}$Li-Cs \cite{PhysRevLett.119.233401} and Yb-Cs \cite{PhysRevA.103.033306}. Additionally, the 780-nm beam helps guide the Li atoms from the Li ODT to the Cs ODT. With the beam, we increase the number of Li transferred by a factor of 2.3.

Prior to the merge, we have $1.9\times10^5$ Li atoms at $\approx$5~$\mu$K and $7.6\times10^5$ Cs atoms at $\approx$1~$\mu$K. After successfully merging the two species, the numbers of Li and Cs atoms change to $7.9\times 10^4$ and $7.3\times10^5$, respectively, while the temperatures of the two species reach equilibrium at 1 $\mu$K. This outcome demonstrates the effectiveness of our merging approach.

\section{sympathetic cooling}\label{sympathetic_cooling}

Evaporative cooling continues after merging the two species. During this stage, the evaporation process involves reducing the optical power of the laser beams for the bichromatic optical dipole trap. Lowering the power of the 1550-nm beams weakens the trap for both species. Conversely, reducing the power of the 780-nm beam weakens the trap for Li but strengthens it for Cs. In the experiment, we observe that the positions of Li and Cs atoms are highly sensitive to the alignment of the 780-nm beam. When the beam is aimed slightly below the two species, it shifts Cs atoms upward and Li atoms downward, bringing the two species closer. However, a slight misalignment of the beam can have the opposite effect, deteriorating the overlap. During the initial 500 ms of this evaporation stage, we gradually turn off the beam to simplify the trap configuration.

After extinguishing the 780-nm beam, the trap depths and positions of Li and Cs are governed by the interplay between the 1550-nm trap, the gravitational pull, and the magnetic-field gradient. We adjust these trap parameters to ensure that Li and Cs maintain similar trap depths and positions as far into the evaporation process as possible; specific details are provided in the Appendix. As evaporation progresses and the sizes of the atom clouds shrink, maintaining sufficient overlap becomes challenging. To overcome this, we reactivate the 780-nm beam to enhance the overlap. Simultaneously, we adjust the field gradient to -2.7~G/cm. This change results in a stronger downward magnetic force on Li than on Cs, drawing Li closer to Cs. At the end of the evaporation, the two species are confined in a trap with trapping frequencies of $(\omega_x,\omega_y,\omega_z)$ = $2 \pi \times(110, 63, 535)$~Hz for Li and $2\pi\times(17, 21, 126)$~Hz for Cs.

\begin{figure}[t]
\centering
\includegraphics[width=0.5\textwidth]{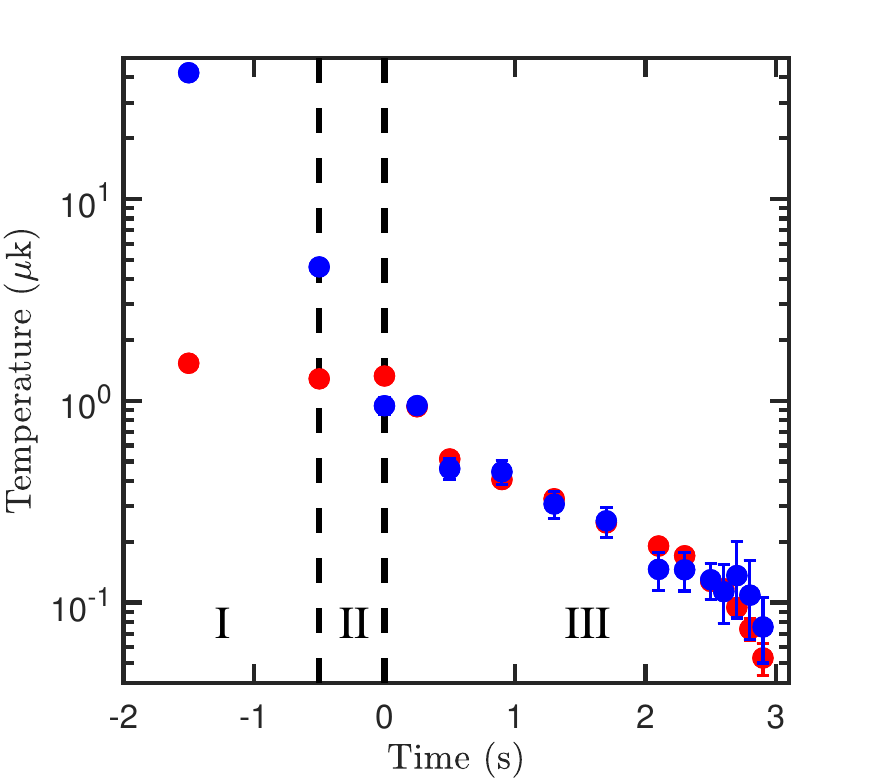}
\caption{Temperatures of Li (blue circles) and Cs (red circles) vs cooling time. The figure is divided into three parts: before the merge (I), during the merge (II), and after the merge (III). Immediately before the merge, Li and Cs exhibit respective temperatures of 4.5 and 1.2 $\mu$K. Immediately after the merge, the temperature of Li reduces to approximately 1 $\mu$K, thus becoming similar to that of Cs. As evaporation continues, the temperatures of the two species remain similar. We observe the formation of the Li BEC at 2.6 s, and the Cs BEC at 2.7 s. As the experiment progresses, the error in Li temperature measurements increases due to the decreasing number of Li atoms. }
\vspace{-4mm}
\label{fig:temperature vs time}
\end{figure}

In the final phase of the evaporation, the 780- and 1550-nm beams combined create a shallow trap for Cs, making the alignment of the beams critical. A misalignment by a few micrometers in their relative beam position can open the trap. Currently, we rely on passive stabilization methods to suppress any potential drift of the beam; however, implementing an active feedback system in the future could significantly improve the beams' precise positioning \cite{PhysRevA.103.033306}.

Throughout this evaporation stage, the magnetic field is kept at 886 G, which results in intraspecies scattering lengths of 484$a_0$ for Li \cite{CRPhys.12.4} and 123$a_0$ for Cs \cite{PhysRevA.87.032517}. With larger scattering length and higher mean velocity, Li has a higher collision rate than Cs, thereby making the Li atoms  more effective at evaporating energy out of the system.

In our experiments, we observe that the temperatures of the two species closely track each other following the merge, extending all the way to the BEC regime, as shown in Fig.~\ref{fig:temperature vs time}. Once the BEC forms, we determine the temperatures by fitting the thermal wings of the clouds. In this regime, only a few thousand Li atoms remain in the BEC phase. The low signal-to-noise ratio introduces larger fitting errors for Li. Still, despite the Li trap depth remaining nearly constant during the final phase of evaporation, our measurements indicate a steady decline in Li temperature and a consistent growth in the Li BEC fraction.

\begin{figure}[t]
\centering
\includegraphics[width=0.5\textwidth]{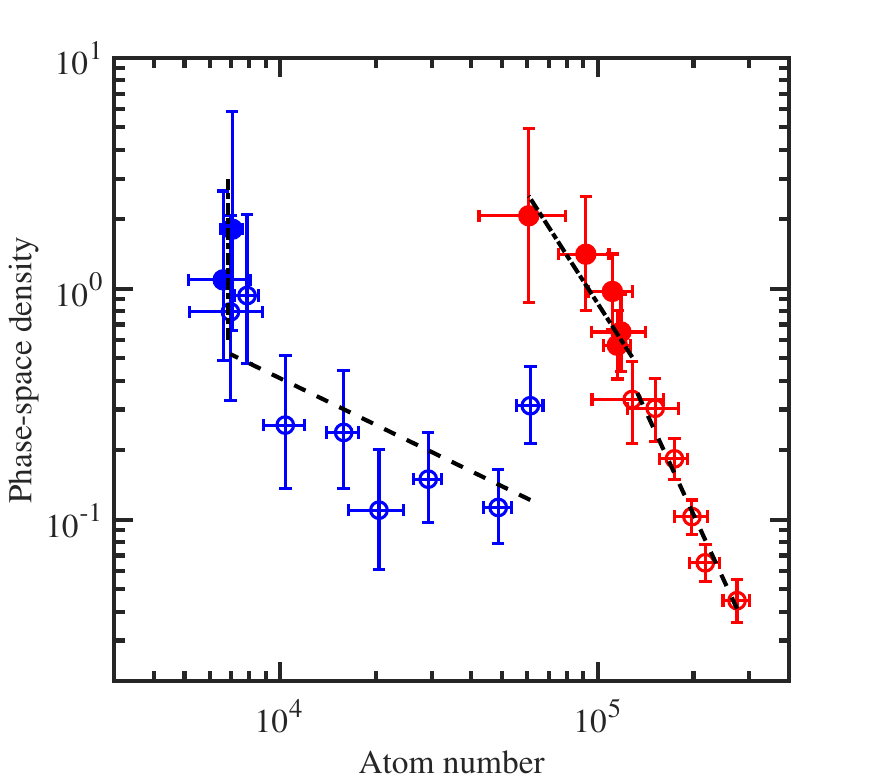}
\caption{Phase-space density vs atom number for Li (blue circles) and Cs (red circles) in the 1550-nm trap. In the initial phase of the evaporation, we observe a Li evaporation efficiency of $\eta_{\rm{Li}} = 0.7$. Meanwhile, the Cs evaporation efficiency, $\eta_{\rm{Cs}}$, is measured at 3.0. Following the reactivation of the 780-nm beam for subsequent evaporation, we observe a sharp increase in Li efficiency alongside a decrease in that of Cs. We use open circles to represent measurements taken before the activation of the 780-nm beam, and solid circles for those taken after the activation. Dashed and dotted-dashed lines serve as guides to the eye for the PSD data collected before and after the beam activation, respectively.}
\label{fig:PSD vs N}
\end{figure}

Figure~\ref{fig:PSD vs N} shows the evaporation trajectories, plotting phase-space density (PSD) against atom number. We use $\eta = -d\ln(\mathrm{PSD})/d \ln(\mathrm{N})$ to evaluate the evaporation efficiency for both Li and Cs. At 500 ms after merging, Li and Cs are co-trapped in the 1550-nm trap. Following this, we gradually reduce the trap depths for both species over the next 1.8 s. It is noteworthy that during this time the two species are subjected to similar trap depths (see Appendix). We measure a significantly lower evaporation efficiency for Li ($\eta_{\rm{Li}}$ = 0.7) compared to Cs ($\eta_{\rm{Cs}}$ = 3.0). After the reactivation of the 780-nm beam, the trap depth for Cs decreases at a greater rate, while that of Li increases slightly. During this final phase of evaporation, there is a significant surge in the PSD of Li. Notably, in this phase, we did not observe any loss of Li atoms beyond the typical range of fluctuation. At the same time, the evaporation efficiency of Cs ($\eta_{\rm{Cs}}$) slightly decreases to 2.1. We attribute this abrupt increase in $\eta_{\rm{Li}}$ to the sympathetic cooling of Li by Cs, an effect only observable with the 780-nm beam. It is also worth noting that single-species BEC formation is only possible for Cs, while the formation of a Li BEC requires the assistance of Cs through sympathetic cooling in the current experimental condition.

\begin{figure}[t]
	\centering
	\includegraphics[width=0.5\textwidth]{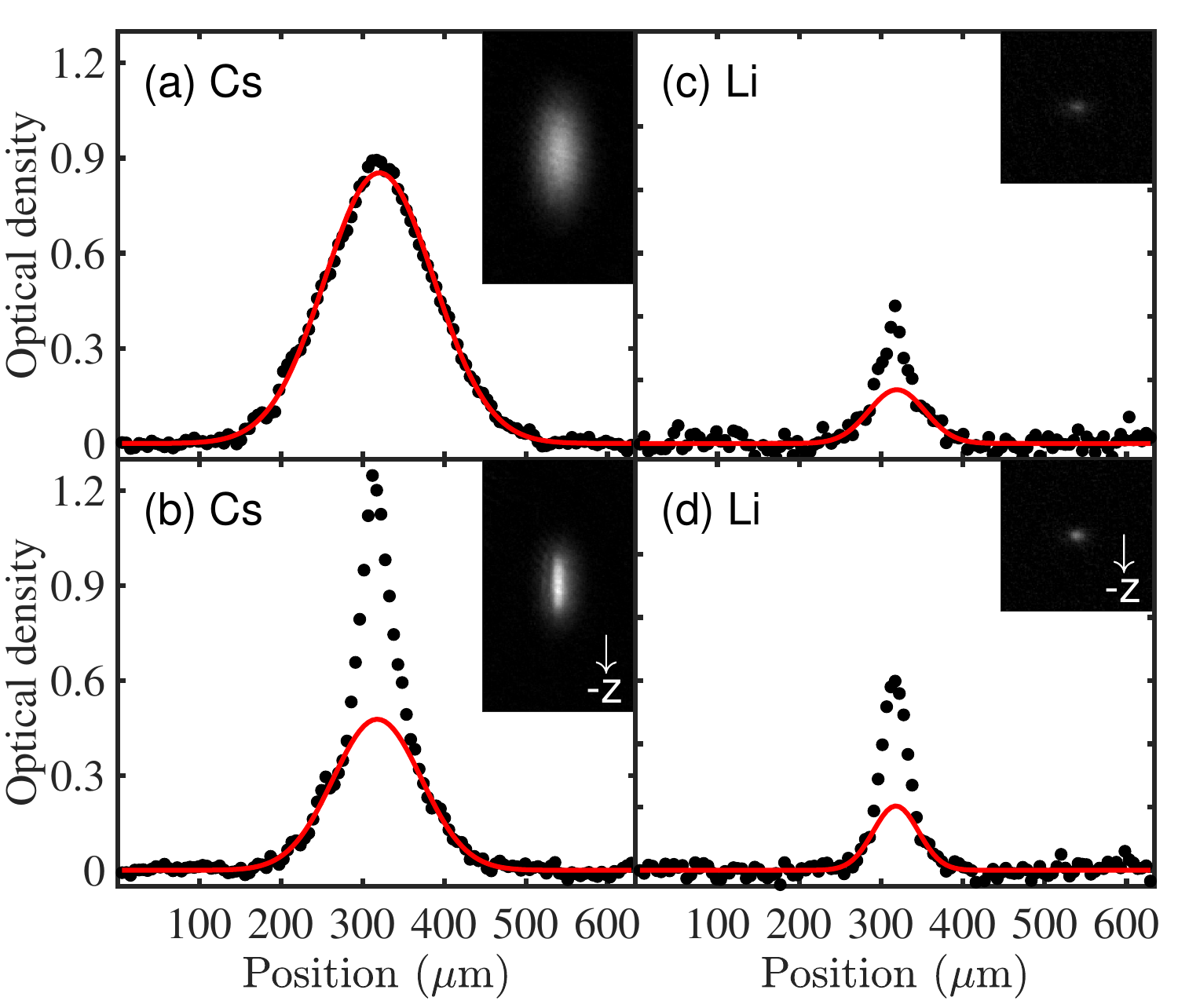}
	\caption{ [(a), (b)] Cs and [(c), (d)] Li center-sliced column densities derived from the time-of-flight side view images, as shown in the insets. The onset of Cs Bose-Einstein condensation is observed at $T=94$~nK, the temperature at which the inset images in (a) and (c) are taken. At a lower temperature of $T=62$~nK, where (b) and (d) are taken, the system achieves dual-species condensation. The red solid lines represent Gaussian fits to the wings of the distributions. The times of flight for Li and Cs are 2.5 and 35 ms, respectively.}
	\label{fig:DualBECimage}
\end{figure}

\section{Dual-Species Bose-Einstein condensation}\label{BEC}
As the evaporation progresses, we observe that Li reaches Bose-Einstein condensation before Cs (see Fig.~\ref{fig:DualBECimage}), with respective transition temperatures of 115 and 95 nK. The Li BEC we observe typically contains $6.0\times 10^3$ Li atoms. Simultaneously, we can create a Cs BEC, consisting of approximately $1.5\times10^4$ Cs atoms. 

The miscibility of quantum gas mixtures plays a crucial role in their potential applications. In a homogeneous system with similar numbers for two species, the miscibility is determined by the balance between intraspecies and interspecies interactions. In this context, the miscible-immiscible transition can be characterized by the miscibility parameter $\Delta = g_{11}g_{22}/g_{12}^2-1$, where the coupling constant is defined as $g_{ij} = 2\pi\hbar^2a_{ij}(m_i+m_j)/m_i m_j$. However, in practice, factors such as density imbalance and variations in trapping potentials can also influence the miscibility.

\begin{figure}[t]
	\centering
	\includegraphics[width=0.5\textwidth]{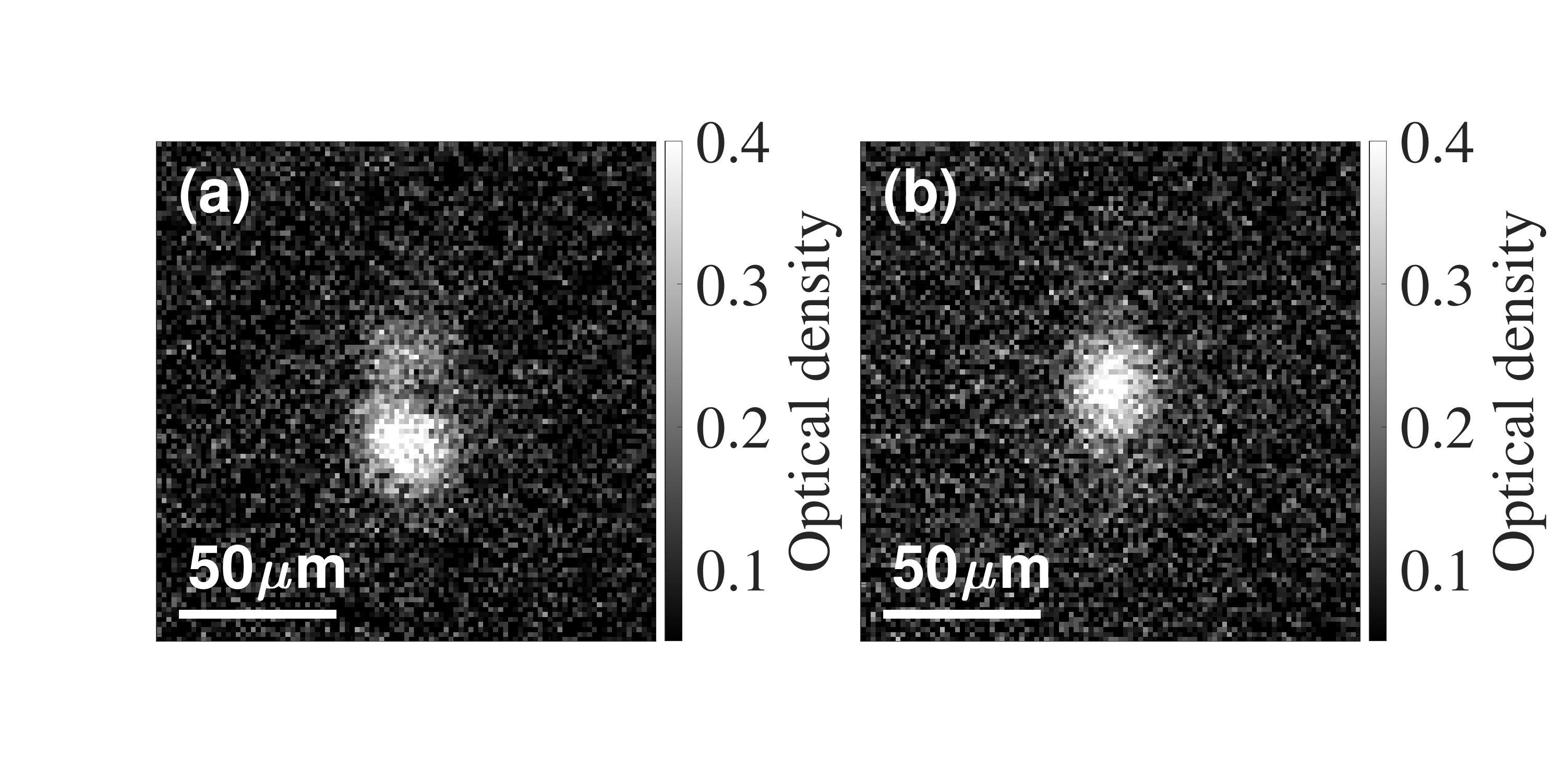}
	\caption{{\it In situ} images (top view) of Li taken at the lowest Li temperatures with (a) and without (b) a Cs BEC. In the presence of Cs BEC, the Li BEC is repelled out of the trap center. Both Li BECs in the images contain approximately 6,000 Li atoms and are at a temperature of 75~nK. The Li trap frequencies here are $(\omega_x,\omega_y,\omega_z)$ = $2 \pi(110, 63, 535)$~Hz.  }
	\label{fig:Li in-situ images}
\end{figure}

During the final stage of evaporation, the bias magnetic field is maintained at 886 G. We derive the interspecies scattering length $a_{\rm{LiCs}}=248a_0$ at this field, according to Ref. \cite{PhysRevA.106.023317}. Given  $a_{\rm{Cs}} = 123a_0$ and $a_{\rm{Li}} = 484a_0$, the miscibility parameter $\Delta$ equals to -0.8, indicating an immiscible phase. To further understand the miscibility of the dual-species BEC of Li and Cs, we solve the full 3D coupled Gross-Pitaevskii equations. The numerical results confirm an immiscible phase given the condensate densities and trap parameters. In the experiment, we observe that the presence of a Cs BEC repels the Li BEC from the trap center (see Fig.~\ref{fig:Li in-situ images}). While we have observed that reducing the temperature of Cs leads to an increase in the size of the Li BEC, further research is required to understand to what extent miscibility impacts the sympathetic cooling process.

\section{Conclusion}\label{conclusion}
We have successfully produced dual-species Bose-Einstein condensates of lithium and cesium. These condensates are created under conditions in which a bias field of 886 G is applied. The bias field results in the intraspecies and interspecies interactions that are all repulsive, and the values of scattering lengths lead to an immiscible phase for the condensates. Still, the Li Feshbach resonance near 894 G and the broad Cs resonance at 787 G could potentially offer sufficient tunability in the intraspecies scattering lengths for studying overlapping superfluids. Specifically, at B = 892 G, the miscibility parameter $\Delta$ is 0.5, a positive but not particularly large value. We can make the value of $\Delta$ even larger by moving the field closer to the Li resonance at 894 G, but the large scattering length of Li at this field may lead to excessive three-body losses, imposing a challenge for future applications. Nevertheless, the $^7$Li-$^{133}$Cs mixtures could still be useful in studies of solitons and polarons. Finally, the Li-Cs system could serve as a promising starting point for creating ground-state LiCs molecules and for investigating Efimov trimers with extremely large and small mass ratios.

As for the impact of miscibility on our proposed experiments, we anticipate its influence on the study of Efimov physics to be minimal since Efimov measurements are typically conducted above the BEC transitions. In our system, the formation of a Cs BEC generates a high-density region at the trap center, which consequently repels Li atoms from this area. By keeping Cs just above quantum degeneracy, we can enhance the spatial overlap of the two species. Therefore, within the context of polaron experiments, it seems more advantageous to study Bose polarons with heavier impurities rather than lighter ones. The Li-Cs Feshbach resonance at 732 G could potentially be used for this type of study. Finally, while the immiscible phase of the dual-species BEC could possibly decrease the conversion rate of Li-Cs Feshbach molecules, the extent of this impact warrants further investigation.

\vspace{-1mm}
\begin{acknowledgements}
This work was supported by the Ministry of Education and the National Science and Technology Council of Taiwan. S.T. acknowledges additional support from the Multidisciplinary and Competitive Programs for Higher Education Sprout Project.
\end{acknowledgements}
%\nocite{*}

\begin{figure}[t]
	\centering
	\includegraphics[width=0.45\textwidth]{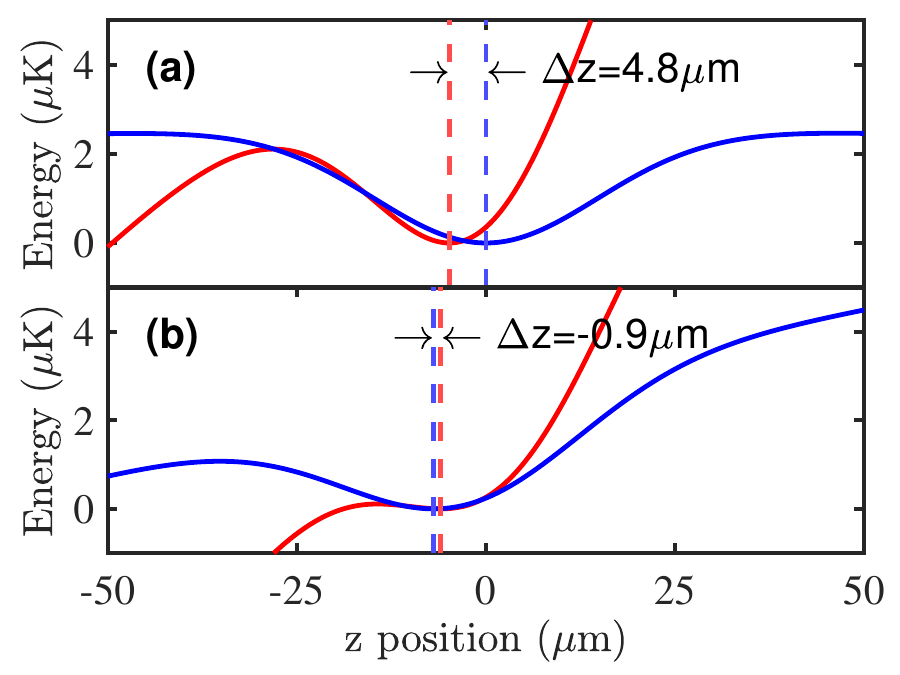}
	\caption{Trapping potentials for Li (blue) and Cs (red) at evaporation times $t$ = 500 ms (a) and $t$ = 2900 ms (b). The referenced time corresponds to that shown in Fig.~\ref{fig:expt_proc}. The dashed lines indicate the positions of the trap minimum.}
	\label{fig:trapping potential}
\end{figure}

\begin{figure}[b]
	\centering
	\includegraphics[width=0.45\textwidth]{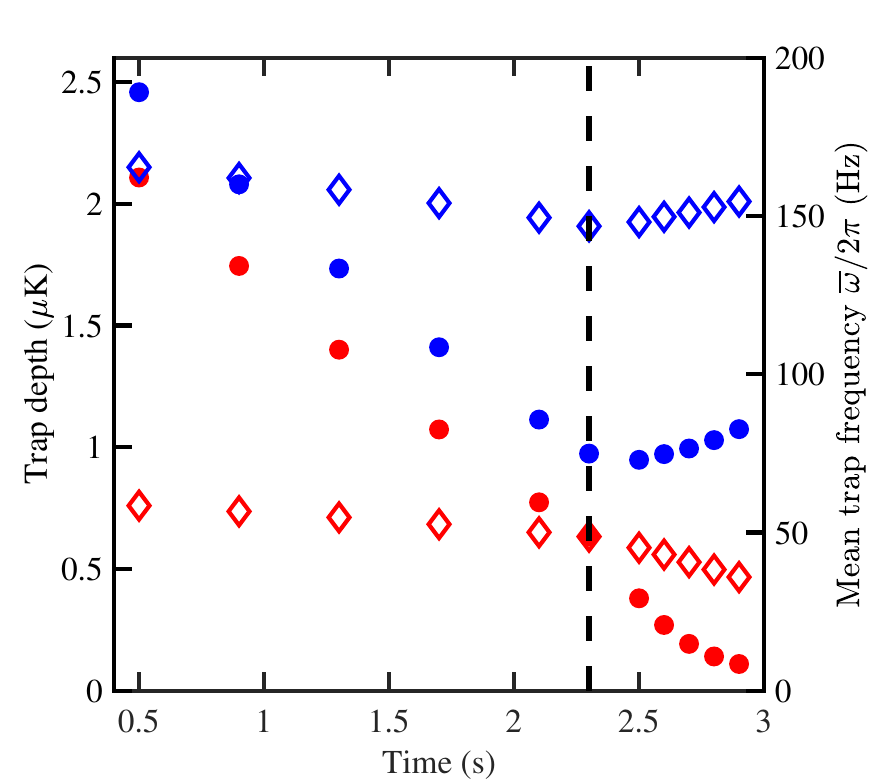}
	\caption{Theoretical values for trap depths (circles) and mean trapping frequencies (diamonds) of Li (blue) and Cs (red), plotted as a function of evaporation time. The dashed line represents the moment when the 780-nm beam is reactivated. }
	\label{fig:trapping_params}
\end{figure}
\vspace{-5mm}
\appendix*
\section{}

The evaporation of Li and Cs after merging takes place within a trap that integrates the effects of multiple components. These include the 1550-nm trap, the 780-nm beam, the magnetic-field gradient, and the gravitational pull. To accurately model the trap, it is necessary to account for each of these elements.

In our calculation, the optical potential is calculated as
\begin{equation}
	\begin{split}
	\label{trapping potential}
	U_D(\mathbf{r}) =  &-\frac{\pi{}c^2}{2}\frac{\Gamma_{D_1}}{\omega_{D_1}^3}(\frac{1}{\omega_{D_1}-\omega}+\frac{1}{\omega_{D_1}+\omega})I(\mathbf{r})\\
	&-\pi c^2\frac{\Gamma_{D_2}}{\omega_{D_2}^3}(\frac{1}{\omega_{D_2}-\omega}+\frac{1}{\omega_{D_2}+\omega})I(\mathbf{r}),
	\end{split}
\end{equation}
where $\Gamma_{D1}$ ($\Gamma_{D2}$) is the linewidth of the $D_1 (D_2)$ line, $\omega_{D_1} (\omega_{D_2})$ is the transition frequency of the atomic $D_1 (D_2)$ line, $\omega$ is the optical frequency of the dipole trap, and $I(\mathbf{r})$ is the intensity profile of the trap. We assume a Gaussian profile for all the laser beams. For a beam propagating along the $y$ axis, the beam profile takes the form 
\begin{equation}
	I(\mathbf{r}) = \frac{2P_0}{\pi{}w_xw_z}e^{-2x^2/w_x^2-2z^2/w_z^2},
\end{equation}
where $w_x$ and $w_z$ are the beam waists, and $P_0$ is the power of the beam. 

It is straightforward to write the gravitational potential:
\begin{equation}
	U_g = - mgz.
\end{equation}
In addition, the potential from the magnetic-field gradient is given by 
\begin{equation}
	U_B = \frac{\partial{E}}{\partial{B}}B^{\prime}z,
\end{equation}
where $E$ is the Zeeman level energy and $B^{\prime}$ is the magnetic gradient. The respective values of $\partial E/ \partial B$ are $-1.18h$ MHz/G for Cs and $-1.33h$ MHz/G for Li at $B$ = 886 G.
 
The potential of the trap is notably sensitive to the 780-nm beam. To ensure stability during evaporation, we ramp down the power of the beam at the early phase of the evaporation. In its absence, the resulting trap depths for both Li and Cs species become comparable, and the separation of the trap minima for the two species amounts to 4.7~$\mu$m, as shown in Fig.~\ref{fig:trapping potential}(a). Subsequently, evaporation is carried out by lowering the 1550-nm trap depth. In the final evaporation phase, we reactivate the 780-nm beam, increasing its power from 0 to 32 mW and adjusting the magnetic gradient from 3.2 to -2.7 G/cm. The resulting potential is depicted in Fig.~\ref{fig:trapping potential}(b). The inversion of the magnetic-field gradient pushes both species towards the direction of gravity, with the effect being more pronounced for Li. This force brings Li closer to Cs. This final potential yields a separation of 0.9~$\mu$m between Li and Cs. Lastly, Fig.~\ref{fig:trapping_params} presents the theoretical values for the trap depths and mean trapping frequencies of both Li and Cs, depicted as a function of the evaporation time.

\end{document}